\documentstyle[prd,twocolumn,aps,epsf]{revtex}
\newcommand{\singlefig}[2]{
\begin{center}
\begin{minipage}{#1}
\epsfxsize=#1
\epsffile{#2}
\end{minipage}
\end{center}}
\newcommand{\gsim}{\mbox{\raisebox{-1.ex}{$\stackrel
     {\textstyle>}{\textstyle\sim}$}}}

\newcommand{\beq}{\begin{equation}}
\newcommand{\eeq}{\end{equation}}
\newcommand{\beqa}{\begin{eqnarray}}
\newcommand{\eeqa}{\end{eqnarray}}
\newcommand{\bea}{\begin{array}}
\newcommand{\ena}{\end{array}}
\begin{document}
\draft
\title{Internal structure of Skyrme black hole}
\author{Takashi Tamaki
\thanks{electronic mail:tamaki@gravity.phys.waseda.ac.jp}
and Kei-ichi Maeda\thanks{electronic
mail:maeda@gravity.phys.waseda.ac.jp}
}
\address{Department of Physics, Waseda University,
Ohkubo, Shinjuku, Tokyo 169-8555, Japan}
\author{Takashi Torii
\thanks{electronic mail:torii@resceu.s.u-tokyo.ac.jp}
}
\address{Research Center for the Early Universe, 
University of Tokyo, Hongo, Bunkyo, Tokyo 113-0033, Japan
\\ and \\
Advanced Research Institute for Science and Engineering,
Waseda University, Ohkubo, 
Shinjuku, Tokyo 169-8555, Japan}
\date{\today}
\maketitle
%------------------------------
\begin{abstract}
We consider the internal structure of the Skyrme black hole 
under a static and spherically symmetric ansatz. 
We concentrate on solutions with the node number one and 
with the ``winding" number zero, where there exist two solutions 
for each horizon radius; one solution is stable and the other 
is unstable against linear perturbation. 
We find that a generic solution
exhibits an oscillating behavior near the sigularity, as 
similar to a 
solution in the Einstein-Yang-Mills (EYM) system, 
independently to stability of the solution. 
Comparing it with 
that in the EYM system, this oscillation becomes mild 
because of the mass term of the Skyrme field. We also find 
Schwarzschild-like exceptional solutions where no oscillating 
behavior is seen. Contrary to the EYM system where there is 
one such solution branch if the node number is fixed, 
there are two branches corresponding to 
the stable and the unstable ones. 
\end{abstract}

%%%%%%%%%%%%%%%%
\section{Introduction}
%%%%%%%%%%%%%%%%
After the discovery of the Bartnik-Mckinnon solution 
in the SU(2) Einstein-Yang-Mills (EYM) system\cite{Bartnik}, 
many particle-like and black hole solutions were found 
and argued in extended systems such as the EYM-Higgs (EYMH) 
system and the EYM-dilaton (EYMD) system\cite{Gal'tsov}. 
They are important as counterexamples of 
the black hole no hair conjecture\cite{Bizon}. 
Moreover, recent investigations show that the internal 
structure of a black hole in the SU(2) EYM system 
exhibits interesting behaviors which are not found 
in the Kerr-Newman black holes\cite{Donets,Maison}. 
A generic solution  in the EYM system exhibits an infinitely 
oscillating behavior 
near the sigularity, which can be understood by reducing the equations 
to a simplified dynamical system. Though this oscillation becomes 
more violent towards the singularity, it is not chaotic. 
This result is extended to other systems such as the 
EYMH system\cite{Maison,Higgs} and the EYMD system\cite{dilaton}. 
In these cases, a scalar field plays a crucial role in preventing 
such behaviors. 

At present, the internal structure of the Einstein-Skyrme (ES) system 
has not been analyzed by previous researches\cite{LM,Skyrme,Bizon2}. 
This system, however, may provide us 
an interesting example, since the mass term of the Skyrme field 
can give  nontrivial effects when we reduce this system to 
the simplified dynamical system\cite{private}. 
If we see the ES system as an effective theory of the 
EYMH system, it is plausible that the 
oscillating behavior may be stopped in spite of 
the absence of the scalar field. 
Moreover, we are interested in its relation to the stability of 
black holes. As a common feature to the EYMH system, 
there are two black hole solutions in the ES system if we fix 
the horizon radius, the node number 
and the ``winding" number\cite{Bizon2,Torii}. 
This result was also extended to 
Brans-Dicke theory\cite{Tamaki}. One of the solutions
is similar to the Schwarzschild 
black hole and stable, and the other to the black hole in the 
EYM system 
and unstable. 
We want to know how the difference of these
solutions are reflected in their internal structures. 
After describing our model, we investigate these features and 
compare them with those in the EYM system. 
Throughout this paper, we use the units $c=\hbar =1$. 

%%%%%%%%%%%%%%%%
\section{Model and Basic Equations}
%%%%%%%%%%%%%%%%

We start with the following action. 
%%%%%%%%%%%%%%%%%%%
\begin{eqnarray}
S=\int d^{4}x\sqrt{-g}\left[\frac{1}{2\kappa^{2}}R
+L_{m}\right]\ ,
\label{eq:1}
\end{eqnarray}
%%%%%%%%%%%%%%%%%%%
where $\kappa^{2} := 8\pi G$ with $G$ being Newton's
gravitational constant. 
$L_{m}$ is the Lagrangian of the matter field. 
We choose the Skyrme  field $L_{m}$, which is
$SU(2)\times SU(2)$ invariant and is given as\cite{Skyrme-orig}
%%%%%%%%%%%%%%%
\begin{eqnarray}
L_{m}=-\frac{1}{32 g_{s}^{2}}{\rm Tr}\mbox{\boldmath $F$}^{2}-
\frac{f_{s}^{2}}{4}{\rm Tr}\mbox{\boldmath $A$}^{2}\ ,
\label{eq:s1}
\end{eqnarray}
%%%%%%%%%%%%%%%
where $f_{s}$ and $g_{s}$ are coupling constants. In our
convention, 
$g_{s}$ is related to the coupling constant of the Yang-Mills field 
$g_{c}$ as $g_{s}=\sqrt{4\pi}g_{c}$.
The ``mass" parameter of the Skyrme field is defined by
$\mu :=f_{s}g_{s}$. $\mbox{\boldmath $F$}$ and 
$\mbox{\boldmath $A$}$ are the  field
strength  and  its potential, respectively. They are  described by the 
$SU(2)$-valued function
$\mbox{\boldmath $U$}$ as
%%%%%%%%%%%%%%%
\begin{eqnarray}
\mbox{\boldmath $F$}= \mbox{\boldmath $A$}\wedge\mbox{\boldmath
$A$}, ~~~
\mbox{\boldmath $A$}=\mbox{\boldmath $U$}^{\dag}
\nabla\mbox{\boldmath $U$}\ .
\label{eq:s2}
\end{eqnarray}
%%%%%%%%%%%%%%%

In this paper, we consider the static and 
spherically symmetric metric, 
%------------------------------
\begin{eqnarray}
ds^{2}=-C(r) e^{2\delta (r)}dt^{2}  
+C(r)^{-1}dr^{2}+r^{2}{d\Omega_2}^{2} \ ,
\label{metric}
\end{eqnarray}
%------------------------------
where $C(r)=:1-2Gm(r)/r$. In this case, we can set 
$\mbox{\boldmath $U$}$, 
%%%%%%%%%%%%%%%%
\begin{eqnarray}
\mbox{\boldmath $U$}(\chi)=\cos \chi (r)+
i\sin \chi (r) \mbox{\boldmath $\sigma$}_{i}\hat{r}^{i}\ ,
\label{eq:s3}
\end{eqnarray}
%%%%%%%%%%%%%%%%
where $\mbox{\boldmath $\sigma$}_{i}$ and $\hat{r}^{i}$ are the Pauli 
spin matrices and a radial normal, respectively. 
The boundary condition for a black hole solution at spatial infinity
is, 
%%%%%%%%%%%%%%%%%%
\begin{eqnarray}
\lim_{r\rightarrow \infty} m = M< \infty,\ \ 
\lim_{r\rightarrow \infty}\delta =0\
\ \label{eq:2.2}.
\end{eqnarray}
%%%%%%%%%%%%%%%%%%
For the existence of a regular event horizon, $r_h$, we have
\begin{eqnarray}
m_{h}:= m(r_h)=\frac{r_{h}}{2G},\ \ 
\delta_{h}:=\delta(r_h)<\infty\ .  \label{eq:12}
\end{eqnarray}
We also require that no singularity exists outside the
horizon, i.e.,
\begin{eqnarray}
m(r)<\frac{r}{2G}\  ~~~~~{\rm  for} ~~~ r>r_{h}. \label{eq:15}
\end{eqnarray}
The boundary condition of a Skyrme field 
for the total field energy to be finite is written generically as 
%%%%%%%%%%%%%%%%
\begin{eqnarray}
\lim_{r\rightarrow \infty}\chi =0\ .    \label{eq:s4}
\end{eqnarray}
%%%%%%%%%%%%%%%%

For our numerical calculation, we introduce the following 
dimensionless variables:
%%%%%%%%
\begin{eqnarray}
\bar{r}=r/r_{h},\ \bar{m}=Gm/r_{h}.
\label{eq:2.3}
\end{eqnarray}
%%%%%%%%
We also define dimensionless parameters as
\begin{eqnarray}
\bar{f}_{s}:= f_{s}/m_{p},\ \ \lambda_{h}=r_{h}/(l_{p}/g_{c}) \ .
\label{eq:6}
\end{eqnarray}
$l_{p} := G^{1/2}$ and $m_{p} := G^{-1/2}$ are the Planck
length and mass defined by Newton's gravitational constant, 
respectively. The basic equations are now
%%%%%%%%%%%%%%%%
\begin{eqnarray}
\frac{d\bar{m}}{d\bar{r}}&=&\frac{1}{\lambda_{h}^2}
\left[BC\left(\frac{d\chi}{d\bar{r}}\right)^{2}
+\frac{A\sin^{2}\chi}{
2\bar{r}^{2}}
\right]   \ ,
\label{eq:sefra1}  \\
%%%%%%%%%%%%%%%%%%%%%%%%%
\frac{  d\delta  }{  d\bar{r} }&=&-
\frac{2B}{\lambda_{h}^{2}\bar{r}} 
\left(  \frac{d\chi}{d\bar{r}}  \right)^{2}
\ ,  \label{eq:sefra2}  \\
%%%%%%%%%%%%%%%%%%%%%%%%%%%
\frac{d^{2}\chi}{d\bar{r}^{2}}&=&
-\frac{1}{2B}
\left(
8\pi\lambda_{h}^{2}\bar{f}_{s}^{2}\bar{r}
+\frac{d\chi}{d\bar{r}}\sin 2\chi
\right)\frac{d\chi}{d\bar{r}}
\nonumber  \\
&&+\frac{1}{C}
\left[
\frac{\sin 2\chi}{2B}
\left(
4\pi\lambda_{h}^{2}\bar{f}_{s}^{2}+\frac{  \sin^{2}\chi  }{
\bar{r}^{2}  }
\right)  \right.
\nonumber  \\
&&+\left. \frac{d\chi}{d\bar{r}}
\left(
\frac{A\sin^{2} \chi}{\lambda_{h}^{2}\bar{r}^{3}}
-\frac{2\bar{m}}{\bar{r}^{2}}
\right)
\right]
\ ,  \label{eq:sefra4}
\end{eqnarray}
%%%%%%%%%%%%%%%%%%%%%
where we use abbreviations as 
%%%%%%%%%%%%%%%%%%%%%
\begin{eqnarray}
A&:=&8\pi\lambda_{h}^{2}\bar{f}_{s}^{2}\bar{r}^{2}+\sin^{2}\chi\ ,
\label{ab1}  \\
B&:=&2\pi\lambda_{h}^{2}\bar{f}_{s}^{2}\bar{r}^{2}+\sin^{2}\chi
\ .      \label{ab2}   
\end{eqnarray}
%%%%%%%%%%%%%%%%
As we can see above, the metric function
$\delta$ is decoupled in the equations and 
calculated by just integrating Eq.~(\ref{eq:sefra2})
after we obtain the solution $\chi(r)$ and $m(r)$.
The square bracket in Eq.~(\ref{eq:sefra4}) must vanish at 
$r=r_h$ for the horizon to be regular. Hence
%%%%%%%%%%%%%%%%%%%%%
%%%%%%%%%%%%%%%%%%%%%
\begin{eqnarray}
\frac{  d\chi  }{  d\bar{r}  }\bigg|_{\bar{r}=1}&=&
-\frac{  \lambda_{h}^{2}\sin 2\chi_{h}(
4\pi\bar{f}_{s}^{2}\lambda_{h}^{2}+\sin^{2}\chi_{h})  }
{2B_{h}(A_{h}\sin^{2}\chi_{h}-\lambda_{h}^{2})  }
\ ,      \label{eq:sseisoku1}   
\end{eqnarray}
%%%%%%%%%%%%%%%%
where $\chi_{h}:= \chi (r_{h})$ is a 
shooting parameter and should be determined iteratively so that 
the boundary conditions (\ref{eq:2.2}) and (\ref{eq:s4}) are 
satisfied.

%%%%%%%%%%%%%%%%%%
\section{Results}
%%%%%%%%%%%%%%%% 

\begin{figure}
\begin{center}
\singlefig{10cm}{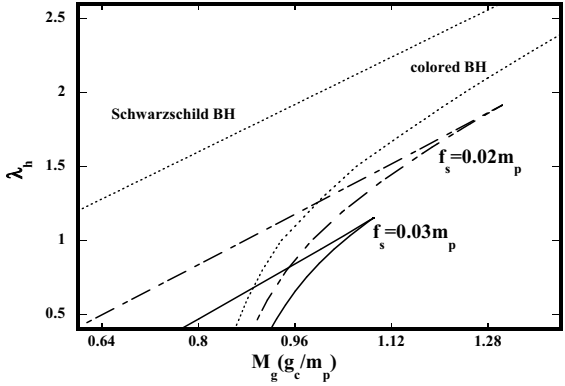}
\caption{$M$-$\lambda_h$ diagram of Skyrme black holes with 
$\bar{f}_{s}=0.02$ and $0.03$. We also show the results of 
the colored and Schwarzschild black holes for reference. 
\label{M-rh} }
\end{center}
\end{figure}

First, we briefly review main properties of the Skyrme black holes 
analyzed in Ref. \cite{Torii}. 
We show the relation between the gravitational mass $M$ 
and the horizon radius $r_{h}$ in Fig. \ref{M-rh}. 
As the colored black hole, i.e., the black hole solution 
in the EYM system, 
these solutions are also characterized 
by the node number. We choose solutions of node number one. 
For Skyrme black holes\cite{LM,Skyrme}, 
the solutions are also characterized by the ``winding"
number defined by\cite{footnote7}
%%%%%%%%%%%%%%%%
\begin{eqnarray}
W_{n}:= \frac{1}{\pi}|\chi_{h}-\chi (\infty)-\sin
(\chi_{h})|\ .
\label{eq:s6}
\end{eqnarray}
%%%%%%%%%%%%%%%%
We consider solutions with the ``winding" number close to one 
(it is $\sim 1$ near the zero horizon limit and $\sim 0.75$ 
near the cusp in Fig.~\ref{M-rh}.) 
with $\bar{f}_{s}=0.02$ and $0.03$ in Fig.~\ref{M-rh}. 
We also show the colored and Schwarzschild black hole cases,
which correspond to $\bar{f}_{s}=0$ limit,
by dotted lines. 
We can find two solution branches for fixed
$\bar{f}_{s}$. These branches merge at some 
critical radius over which a solution disappears. This is because 
the non-trivial structure of the 
non-Abelian field becomes as large as  the scale of the Compton
wavelength ($\sim 1/\mu$). 
That is, beyond this critical horizon radius, a non-trivial
structure is swallowed into the horizon 
resulting in a Schwarzschild spacetime.

Moreover, we should mention that a cusp structure which 
appears at the 
critical radius is a symptom of the stability change in 
catastrophe theory\cite{CAT}. The massive branch is unstable 
while the other branch is stable\cite{Torii}. 
If $f_{s}$ becomes small, the massive unstable 
branch approaches the colored black hole. Thus, it is plausible 
that the internal structure of the Skyrme black hole of the unstable 
branch shows similar behavior to the colored black hole 
when the mass of the 
Skyrme field is small. But, how about 
the other branch or how is this altered by the mass of the 
Skyrme field? These questions have not been considered previously.

\begin{figure}
\begin{center}
\singlefig{10cm}{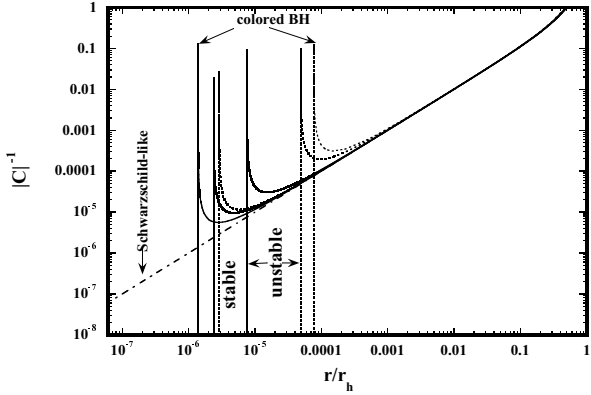}
\caption{$\bar{r}$-$|C|^{-1}$ diagram 
of Skyrme black holes with $\bar{f}_{s}=0.03$ and colored 
black holes. 
We consider both unstable and stable branches with the 
horizon radii $\lambda_{h}=0.7$ (solid lines) and 
$\lambda_{h}=0.9$ (dashed lines). We also show the 
corresponding colored black holes and the exceptional Skyrme 
black hole solution $\lambda_{h}\sim 0.52$ where there is no 
oscillating behavior.  
\label{structure1} }
\end{center}
\end{figure}

We show typical structure ($\bar{r}$-$|C|^{-1}$) found inside 
the Skyrme black hole in Fig.~\ref{structure1} for 
solutions with $\bar{f}_{s}=0.03$ and $\lambda_{h}=0.7$ (solid lines) 
and $0.9$ (dotted lines). For reference, we also show the corresponding 
colored black hole solutions and the exceptional Skyrme black hole 
solution $\lambda_{h}\sim 0.52$ where no oscillating behavior is seen 
(it corresponds to the Schwarzschild-like solution in the analysis below.). 
For field variables, we choose them to satisfy asymptotically flatness. 
In this figure, we find a steep 
peak (a local maximum) 
at some radius independently of the types of black holes. 
Below the radius, $|C|^{-1}$ exhibits a rapid deacrease to minus ten
to several hundreds.
To calculate this behavior correctly, 
it is necessary to use suitable 
variables as introduced in the Appendix of Ref. \cite{Donets}. 
Similarly to the colored black hole, infinitely oscillating 
behavior appears in the Skyrme black hole case. Thus, the mass term 
of the Skyrme field do not help to avoid this behavior. 
It is the main difference in the
internal structure between the ES system and the EYMH system. 
As for the Skyrme field, $\chi$ approaches some constant value 
toward the center. At first glance, it seems strange that 
$\chi$ stays at an almost constant value while 
$|C|^{-1}$ exhibits violent behavior. 
This is, however, also seen in the colored 
black hole case where its gauge potential $w$ 
takes almost a constant value though its gradient 
$|w'|$ takes a large value in some 
extremely narrow interval\cite{Donets}. Since $\chi$ is related to 
$w$ by $\cos \chi =w$ in the $f_{s}\to 0$ limit, 
our result is consistent with the colored black hole case. 

\begin{figure}[htbp]
\singlefig{10cm}{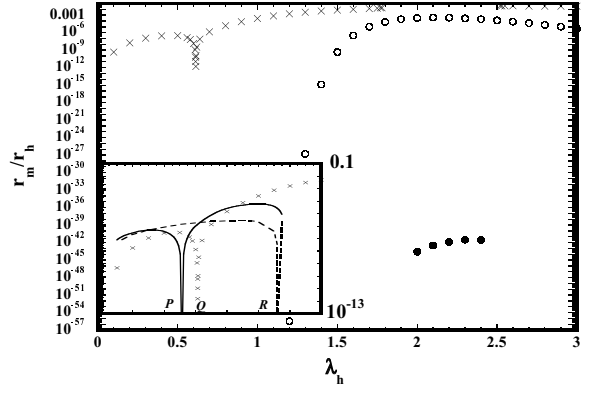}{}
\caption{$\lambda_{h}$-$r_{m}/r_{h}$ diagram of the colored 
black holes. A multiplication sign, a white circle and a 
black circle mean that $|C|^{-1}$ takes a first, a second and a third 
peak, respectively. We also show the unstable and the stable branches of 
the Skyrme black hole in the framed region ($0.1<r_{m}/r_{h}<10^{-13}$ 
and $0<\lambda_{h}<1.4$) which are shown 
by a solid line and by a dotted line, respectively. 
\label{rcrit1}}
\end{figure}

%%%%%%%%%%%
%   comparison  
%%%%%%%%%%%
Comparing the Skyrme black hole with the  colored 
black hole, the first peak of $|C|^{-1}$ appears 
at larger radius for 
$\lambda_{h}=0.7$ while at smaller radius for 
$\lambda_{h}=0.9$. The mass term gives quite complicated effects
on the interior structure like this. To clarify them,
we first show the relation between 
the horizon radius $\lambda_{h}$ and 
$r_{m}/r_{h}$ for colored black holes 
in Fig. \ref{rcrit1}. $r_{m}$ is the radius 
where $|C|^{-1}$ takes a local maximum value. 
Multiplication signs, white circles and a 
dots mean that $|C|^{-1}$ takes a first, a second and a third 
branch of the peaks, respectively. 
At first glance, it may seem strange that the first branch is
missing
in the region $1.8<\lambda_{h}<2.5$. This can be understood as 
follows. For $\lambda_{h}\sim 1.7$, the peak is much milder 
compared with those in Fig.~\ref{structure1}. 
When $\lambda_{h}$ becomes large, the local maximum
disappears by coincidence with the local mimimum.
Similar disappearance occurs around $\lambda_{h}\sim 2.5$.

%%%%%%%%%%%%%%%
%  comparison of r_{m}
%%%%%%%%%%%%%%%
We show the Skyrme black hole case with 
$\bar{f}_{s}=0.03$ in the framed region in Fig.~\ref{rcrit1} 
which corresponds to the magnification of the region 
$0.1<r_{m}/r_{h}<10^{-13}$ and $0<\lambda_{h}<1.4$ since 
the Skyrme black hole solution does 
not exist above $\lambda_{h}\gsim 1.15$ for $\bar{f}_{s}=0.03$. 
The unstable and the stable branches of 
the Skyrme black holes are shown by a solid line 
and by a dotted line, respectively. 
The lines in this figure correspond to the first peak. 
We can see that the oscillating behavior is generic 
and the Skyrme black hole does not have a Cauchy horizon in general 
inspite of the influence by the mass term of the Skyrme field. 
Moreover, these properties do not depend on 
the stability of the black holes. 
This independence must be related to the fact that
although the colored black hole itself is unstable against 
linear perturbation, interior oscillating structure is stable against 
non-linear perturbations\cite{Tsulaia}. 
In other words, this problem belongs to 
the structure of the ordinary differential equations around the center. 

%%%%%%%%%%%%%%%%%
% r_{m} ?
%%%%%%%%%%%%%%%%%
We can find the tendency that $r_m$ for the Skyrme black hole is
larger than that of the colored black hole on average, and that
the amplitude of the oscillation is milder for the Skyrme black hole
than that for the colored black hole. We can 
not say, however, definite things about these properties since
there are exceptional solutions, which make the comparison complicated.

%???????????????????????
%  exceptional cases  
%%%%%%%%%%%%%%%%%%%%%%%%
We finally mention these exceptional cases, where
no oscillating behavior is seen. 
The candidates are the solutions corresponding to 
the points $P$, $Q$ and $R$ in Fig.~\ref{rcrit1}. 
Three types of behaviors are considered 
assuming that variables are expressed by 
power series near the center\cite{Donets,Maison}, 
i.e., Schwarzschild-like, Reissner-Nortstr\"om (RN)-like 
and imaginary charged RN-like behaviors. 
By expanding the field variables  
%%%%%%%%%%%%%%%%%%%%%
\begin{eqnarray}
w=\sum_{n=0}^{\infty}w_{n}\bar{r}^{n},\ \ \ 
\bar{m}=\sum_{n=-1}^{\infty}\bar{m}_{n}\bar{r}^{n}
\ ,      \label{eq:expand}   
\end{eqnarray}
%%%%%%%%%%%%%%%%
where $w:=\cos \chi$, we can summerize these behaviors: 

(i) Schwarzschild-like case ($\bar{m}_{-1}=0$). 
In this case, we can express the above expansion in the 
following form, 
%%%%%%%%%%%%%%%%%%%%%
\begin{eqnarray}
w&=&-1+b\bar{r}^{2}+O(\bar{r}^{4})
\\
\bar{m}&=&\bar{m}_{0}-\frac{4}{\lambda_{h}^{2}}\bar{m}_{0}b(\pi 
\lambda_{h}^{2}\bar{f}_{s}^{2}+b)\bar{r}^{2} 
\nonumber  \\
&&+\frac{2b}{\lambda_{h}^{2}}(2\pi\lambda_{h}^{2}\bar{f}_{s}^{2}+b)\bar{r}^{3}
+O(\bar{r}^{4}).  
\label{eq:Sch-like}   
\end{eqnarray}
%%%%%%%%%%%%%%%%
This is quite consistent with the previous result in Ref. \cite{Donets} 
since it attributes to the EYM case in the $\bar{f}_{s}\to 0$ limit. 
The important thing is that $b$ is {\it not} a free parameter for 
$\bar{f}_{s}\neq 0$ since we 
encounter a relation $\bar{f}_{s}b =0$ in expanding Eq.(\ref{eq:sefra4}). 
Thus, we can find that $w\equiv -1$ and $\bar{m}\equiv\bar{m}_{0}$. 
This means that there is not a Schwarzschild-like solution or that 
$w$ approaches $-1$ faster than $\bar{r}^{n}$ ($n=1,2,3 \cdots$) in 
$\bar{r}_{h}\to 0$. As we see below, 
our numerical analysis support the latter. 

(ii) RN-like case. Below, $w_{0}\neq \pm 1$ is assumed. 
This case contains three free parameters ($w_{0}$, $c$ and $\bar{m}_{0}$) as 
%%%%%%%%%%%%%%%%%%%%%
\begin{eqnarray}
w&=&w_{0}-
\frac{w_0}{2V_{0}}\bar{r}^{2}+\frac{c}{2V_{0}}\bar{r}^{3}
+O(\bar{r}^{4}),
\label{RN-like:chi}\\ 
\bar{m}&=&-\frac{V_{0}^{2}}{2\lambda_{h}^{2}\bar{r}}+
\bar{m}_{0}-4\pi\bar{f}_{s}^{2}V_{0}\bar{r}
\nonumber \\
&&-w_{0}\left(\frac{c}{\lambda_{h}^{2}}+
\frac{\bar{m}_{0}w_{0}\lambda_{h}^{2}}{V_{0}^{2}}\right)
\bar{r}^{2}+O(\bar{r}^{3})\ .   
\label{RN-like:mass}   
\end{eqnarray}
%%%%%%%%%%%%%%%%
where $V_{0}:=w_{0}^{2}-1$. 

(iii) Imaginary charged RN-like case. This case contains 
one free parameter 
$w_{0}$ as 
%%%%%%%%%%%%%%%%%%%%%
\begin{eqnarray}
w&=&w_{0}\pm\lambda_{h}\bar{r}-
\frac{w_0}{2V_{0}}\bar{r}^{2}
+O(\bar{r}^{3}), \label{imRN-like:chi}\\ 
\bar{m}&=&\frac{V_{0}^{2}}{2\lambda_{h}^{2}\bar{r}}\pm
\frac{2w_{0}V_{0}}{\lambda_{h}}
+O(\bar{r})\ .   
\label{imRN-like:mass}   
\end{eqnarray}
%%%%%%%%%%%%%%%%

\begin{figure}[htbp]
\singlefig{10cm}{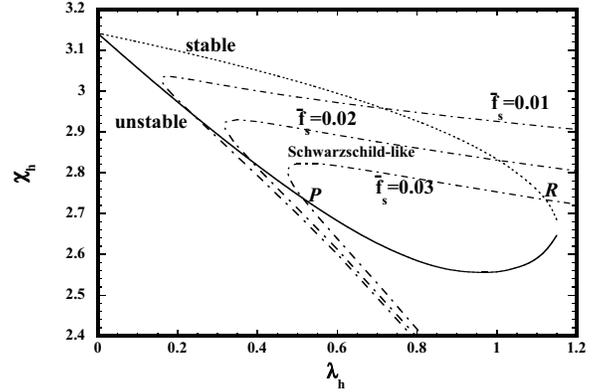}{}
\caption{$\lambda_{h}$-$\chi_{h}$ diagram of the 
Skyrme black holes. $\chi_{h}$ to satisfy the asymptotically 
flatness for $\bar{f}_{s}=0.03$ is shown by a dotted line 
(the stable branch) and 
by a solid line (the unstable branch) and $\chi_{h}$ 
to satisfy the Schwarzschild-like condition for $\bar{f}_{s}=0.01$, 
$0.02$ and $0.03$ 
are shown by dot-dashed lines. 
\label{Schlike}}
\end{figure}

In order to search these exceptioanl cases, 
we have to integrate the basic equations 
toward the center for each Skyrme black hole solution
varying the horizon radius. There is, however,
another approach. We fix the horizon radius and
search the suitable value of $\chi_{h}$ which 
satisfy the above conditions. This $\chi_{h}$ does
not coincide with the value of the Skyrme black hole solution
in general
which satisfy the asymptotically flat condition. 
However, if it coincides, such solution is the 
exceptional one. 
This method was adopted in Refs.~\cite{Donets,Maison} 
and it was found that the point $Q$  
is the exceptional case (i). 
In Fig.~\ref{Schlike},
we exhibit a relation between the horizon radius 
$\lambda_{h}$ and $\chi_{h}$ which satisfies 
the Schwarzschild-like condition (i) for $\bar{f}_{s}=0.01$, 
$0.02$ and $0.03$ (dot-dashed lines). 
We choose $\chi_{h}$ to satisfy $\chi \to \pi$ at $\bar{r}\to 0$. 
We can find two solutions for each horizon radius if the parameter 
$\bar{f}_{s}$ is fixed. Turning back of solutions occurs 
at some critical radius ($\lambda_{h}\sim 0.16$, $0.32$, $0.48$ 
for $\bar{f}_{s}=0.01$, $0.02$ and $0.03$, respectively.) 
below which a solution disappears. 
In the $\bar{f}_{s}\to 0$ limit, this turning back disappears and 
the solution of larger $\chi_{h}$ coincides with exact Schwarzschild 
solution (i.e., $\chi=\pi$). This is consistent with the colored black 
hole case examined in Ref.\cite{Donets,Maison}. 
%%%%%%%%%%%%

We also plot $\chi_{h}$ of the Skyrme black hole solutions for 
$\bar{f}_{s}=0.03$ (i.e., $\chi_{h}$ satisfies asymptotically flatness.). 
We can find the intersections of these solutions 
with the Schwarzschild-like 
solutions. It means that there {\it are} exceptional solutions 
corresponding to 
the points $P$ and $R$ ($\lambda_{h}\sim 0.52,\ \ 1.13$).

%%%%%%%%%%%%%%%%
\section{Conclusion}
%%%%%%%%%%%%%%%%

We examined internal structure of the Skyrme black hole and compared 
it with the colored black hole case. 
Although the gravitational structures and the thermodynamical
properties of the Skyrme black hole are
similar to those of the sphaleron black hole 
in the EYMH system, internal 
structure is  quite different  qualitatively
from the sphaleron black hole because of the 
absence of the scalar field. 
Similarly to the colored black hole, 
the mass function of the Skyrme black hole 
exhibits oscillating behavior. 
The amplitude of the oscillations of the 
metric function tends to become 
smaller than that of the colored black hole due  
to the mass of the Skyrme field. It is difficult, however, to 
conclude that it is a generic feature because of the 
existence of the exceptional solutions where oscillating 
behavior does not happen. We also find exceptional Schwarzschild-like 
solutions which exist both for stable and unstable branches. 

%%%%%%%%%%%%%%%%
\section*{Acknowledgments}
%%%%%%%%%%%%%%%%

Special thanks to Dmitri. V. Gal'tsov and George Lavrelashvili 
for useful discussions. This work was supported by a JSPS Grant-in-Aid 
(No. 106613), and by the Waseda University Grant for Special 
Research Projects.
%%%%%%%%%%%%%%%%%%%%%%%%%%%%%%%%%%%%%%%

%%%%%%%%%%%%%%%%%%%%
%%%%%%%%%%%%%%%%%%%%
%%%%%%%%%%%%%%%%%%%%
%%%%%%%%%%%%%%%%%%%%%%%%
%%%%%%%%%%%%%%%%%%%%%%%%

\begin{thebibliography}{99}
\bibitem{footnote7}
For a particle-like solution (Skyrmion), 
the value of $\chi$ at the origin must be $\pi n$, where $n$ is an 
integer and $|n|$ denotes the winding number of the Skyrmion. In the 
case of the black hole solution, it is topologically trivial. But $W_
{n}$ 
defined by (\ref{eq:s6}) is close to $n$, so we shall also call it the 
``winding" number.
\end{thebibliography}
\end{document}